\documentstyle[12pt]{article}
\def\be{\begin{equation}}
\def\ee{\end{equation}}
\def\ben{$$}
\def\een{$$}
\def\ba{\begin{array}{c}}
\def\ea{\end{array}}
\def\p{\partial}
\begin{document}

\titlepage
\vspace*{2cm}

 \begin{center}{\Large \bf
Re-establishing supersymmetry between harmonic oscillators in $D
\neq 1$ dimensions
  }\end{center}

\vspace{10mm}

 \begin{center}
Miloslav Znojil

 \vspace{3mm}

\'{U}stav jadern\'e fyziky AV \v{C}R, 250 68 \v{R}e\v{z}, Czech
Republic\footnote{e-mail: znojil@ujf.cas.cz}

\end{center}

\vspace{5mm}

\section*{Abstract}

We offer a plausible resolution of the paradox (first formulated
by Jevicki and Rodriguez in Phys. Lett. B 146, 55 (1984)) that the
two shifted  harmonic oscillator potentials $ V(q)=q^2+{G}/{q^2}+
const$  may, in spite of their exact solvability in a non-empty
interval of the couplings $G$, become supersymmetric partners if
and only if $G$ vanishes. We show that and how their $G \neq 0$
SUSY may be re-established via a regularization provided by
pseudo-Hermitian quantum mechanics.

\vspace{5mm}

PACS   03.65.Fd; 03.65.Ca; 03.65.Ge; 11.30.Pb; 12.90.Jv

\newpage

\section{Motivation \label{jedna}}

A suitable algebraic background of the theoretical construction of
multiplets which would unify some experimentally observable bosons
with fermions is provided by the graded Lie algebras, so called
superalgebras. In such a setting, an exceptional role is played by
the linear harmonic oscillator in one dimension, $D = 1$. Indeed,
its Hamiltonian may be factorized and, subsequently, shifted to
the left or to the right,
 \ben
 H^{(LHO)}=p^2+q^2 =A\cdot B - 1 = B \cdot A
+ 1, \ \ \ \ \ A = q+ip, \ \ \ B = q-ip
 \een
 \ben
 H_{(L)}=H^{(LHO)}-1=B\cdot A, \ \ \ \ \ H_{(R)}=H^{(LHO)}+1=A\cdot
 B.
 \een
According to Witten \cite{Witten} one may then introduce a
``super-Hamiltonian" and two ``supercharges",
 \ben
 {
 \cal H}= \left [ \begin{array}{cc} H_{(L)}&0\\ 0&H_{(R)}
 \ea
 \right ]
, \ \ \ \ \ \
 {
 \cal Q}=\left [
 \begin{array}{cc} 0&0\\ A^{}&0
 \ea
 \right ],
 \ \ \ \ \ \
\tilde{\cal Q}=\left [
 \begin{array}{cc}
0& B^{}
\\
0&0 \ea \right ]\
 \een
and notice that they generate a representation of Lie superalgebra
sl(1/1),
 \ben
 \{ {\cal Q},\tilde{\cal Q}
\}={\cal H} , \ \ \ \ \ \ \{ {\cal Q},{\cal Q} \}= \{ \tilde{\cal
Q},\tilde{\cal Q} \}=0, \ \ \ \ \ \ \ \ [ {\cal H},{\cal Q} ]=[
{\cal H},\tilde{\cal Q} ]=0.
  \een
In the language of physics, one can speak about the bosonic {\em
and} fermionic vacuum annihilated by both the supercharges,
 \ben
 \langle q |0,0\rangle = \left [
 \ba
 \exp(-q^2/2)/\sqrt{\pi} \\
 0
 \ea
 \right ], \ \ \ \ \
 {\cal Q}\,|0,0\rangle =
 \tilde{\cal Q}\,|0,0\rangle = 0.
 \een
The ``bosons" themselves may be then introduced as created and/or
annihilated by the first-order differential operators $ {\bf
a}^\dagger\sim B$ and/or $ {\bf a}\sim A$,  respectively. The
parallel creation, annihilation and occupation-number operators
are also easily defined for ``fermions",
 \ben
 {
 \cal F}^\dagger=\left [
 \begin{array}{cc} 0&0\\ 1^{}&0
 \ea
 \right ],
 \ \ \ \ \ \
{\cal F}=\left [
 \begin{array}{cc}
0& 1^{}
\\
0&0 \ea \right ],
 \ \ \ \ \ \
{\cal N}_{\cal F}=\left [
 \begin{array}{cc}
0& 0^{}
\\
0&1 \ea \right ]\ .
 \een
The supercharges become factorized as well, ${\cal Q} \sim {\bf
a}{\cal F}^\dagger$ and $\tilde{\cal Q} \sim {\bf a}^\dagger{\cal
F}$. In terms of the harmonic-oscillator eigenstates $|n\rangle$
the Fock space will be spanned by the states $|n_b,n_f\rangle$
characterized by the presence of $n_b$ bosons and $n_f$ fermions
where $n_f$ is equal to zero or one,
 \ben
|n,0\rangle = \left [
 \ba
 |n\rangle \\
 0
 \ea
 \right ], \ \ \ \ \ \ \ \ \
 |n,1\rangle = \left [
 \ba
 0\\
 |n\rangle
 \ea
 \right ].
 \een
In this way the harmonic oscillator may be understood as a
next-to-trivial superymmetric field theory in a one-dimensional
space-time which unifies the bosons with fermions. More details
may be found, e.g., in the concise review paper \cite{Khare}.

In what follows, we shall analyze what happens if we replace the
one-dimensional $H^{(LHO)}$ by its radial, $D-$dimensional
generalization with any real $D > 2$,
 \ben
  H^{(\alpha)}= -
 \frac{d^2}{dq^2}
+
 \frac{\alpha^2-1/4}{q^2}
+ q^2,  \ \ \ \ \ \ \alpha=(D-2)/2 +\ell, \ \ \ \ \ \ \ell=0, 1,
\ldots.
 \een
We shall be guided by the Witten's supersymmetric quantum
mechanics where the use of the operators $A=\p_q+W$ and
$B=-\p_q+W$ with arbitrary superpotentials $W$ leads to the same
supersymmetric pattern as above, forming the superHamiltonian from
the two partner operators
 \ben
 H_{(L)}=B\cdot A=\hat{p}^2+W^2-W'
 , \ \ \ \ \ \ \ \
 H_{(R)}=A \cdot B=\hat{p}^2+W^2+W'.
 %\label{partner}
 \een
In the spirit of our recent letter \cite{plb} we shall admit that
these operators are pseudo-Hermitian~\cite{Mostafazadeh}.

\section{Bound states in the pseudo-Hermitian setting}

Beyond the elementary harmonic oscillator let us now contemplate a
generalized  superpotential
 \be
 W_{}^{(\gamma)}(r) = r-\frac{\gamma+1/2}{r}\, \ \ \ \ \ \ \ r =
r(x)=x-i\,\varepsilon.
 \label{N}
 \ee
In the other words, we assume that we start from the choice of a
real parameter $\gamma $ and {\em define} the pair $H_{(L,R)}$ of
non-Hermitian operators. Such a recipe gives the partner
Hamiltonians in the above $D-$dimensional harmonic oscillator form
where $D_L \neq D_R$ in general,
 \be
 {H}_{(L)}^{(\gamma)} = {H}_{}^{(\alpha)} -2\gamma-2,
 \ \ \ \ \ \
 {H}_{(R)}^{(\gamma)} = {H}_{}^{(\beta)} -2\gamma, \ \ \ \
 \ \ \ {\alpha}=|\gamma|, \ \ \ \
 \ \ \ \beta=|\gamma+1|\ .
 \label{Mtt}
 \ee
In the light of ref. \cite{ptho} the complexified line of
coordinates $r=x-i\,\varepsilon$ circumvents the singularity in
the origin so that the bound state wavefunctions are regular and
expressible in terms of the Laguerre polynomials,
 \ben
 \psi_{ }^{}(r)
= \frac{N!}{\Gamma(N+\varrho+1)}\cdot
 r^{\varrho+1/2} \exp(-r^2/2) \cdot L_N^{(\varrho)}(r^2).
 %\label{keyf}
  \een
Together with their energies
  \ben
  E_{}^{}
  = E_{N}^{(\varrho)} =4N+2\varrho+2,
\ \ \ \ \ \ \ \
 \varrho= -Q \cdot
\alpha, \ \ \ \ \ \ Q = \pm 1,\ \ \ \ N = 0, 1, \ldots
  %\label{key}
  \een
these states are numbered by the integer $N$ and by the so called
quasi-parity $Q$.

\section{Supersymmetry, pseudo-Hermitian way}

For reasons explained in ref. \cite{ptho} we must assume that
$\gamma \neq 0, 1, 2, \ldots$. Up to that constraint, we may
visualize the above construction as one of the regularizations
recommended in the recent literature \cite{tata}. Here we intend
to summarize and discuss the subject in more detail.

In the first step we notice that the quasi-parity $Q$  coincides
with the ordinary spatial parity $P$ in the limit $\alpha \to
1/2$. In such a limit the basis states are well known (cf.
Appendix A). Once we move to $\alpha \neq 1/2$ we notice that the
quasi-even states $\psi(r) \sim r^{1/2-\alpha}$ still lie below
their quasi-odd complements $\psi(r) \sim r^{1/2+\alpha}$ at any
fixed $N$.

Whenever we choose $\alpha \geq 1$, the limiting transition
$\varepsilon \to 0$ moves the quasi-even solutions out of the
Hilbert space completely. Otherwise, these states remain
normalizable in a way depicted in Figure~1 where the following
ordering is obtained for the $N-$th bunch of the energy levels,
 \be
 E_{(L)}^{(-\alpha)} \ [\equiv a(N)]\leq
 E_{(R)}^{(-\beta)} \ [\equiv b(N)] \leq
 E_{(L)}^{(+\alpha)} \ [\equiv c(N)] \leq
 E_{(R)}^{(+\beta)} \ [\equiv d(N)].
  \label{ordered}
 \ee
This ordering is preserved along all their $\gamma-$dependent
variation. Each of these four curves is just a once broken
straight line but, in our picture, our eyes are guided by an
infinitesimal shift of the levels in such a way that their shape
may be easily followed (one should only recollect that the system
remains undefined at all the integers $\gamma\in I\!\!N$).

In the Figure the physical, Hermitian limiting transition
$\varepsilon \to 0$ has been performed. The general, $\varepsilon
\neq 0$ has been discussed elsewhere \cite{susyho}. We may only
note here that in contrast to the latter and manifestly
non-Hermitian, $\varepsilon \neq 0$ scheme of ref. \cite{susyho},
all our present states belong to the Hilbert space of the ordinary
quantum mechanics. Thus, our new scheme may be interpreted as a
result of a pseudo-Hermitian regularization recipe studied, in
more detail, elsewhere~\cite{Eck}  (cf. also Appendix B for some
more details).

The inspection of Figure 1 reveals a certain generalized
supersymmetry (SUSY) where
the standard requirements of quadratic integrability tolerate the
quasi-odd levels at all $\gamma$ but confine the existence of the
levels $a(n)$ to the very short interval of $\gamma \in (-1,1)$
and the existence of the levels $b(n)$ to the interval of $\gamma
\in (-2,0)$.
As a consequence,
one has to distinguish between the
following five mutually significantly different regimes.

\begin{enumerate}

\item ``Far left"
with $\gamma \in (-\infty,-2)$ and with the complete degeneracy
 \ben
E_{(L)}^{(+\alpha)} \ [\equiv c(N)] = E_{(R)}^{(+\beta)} \ [\equiv
d(N)]
 \een
where the Witten's index vanishes \cite{plb} and where SUSY itself
is broken because the ground state energy remains positive. All
the spectrum is equidistant.

\item ``Near left"
with $\gamma \in (-2,-1)$ and with the mere partial degeneracy
which survives from the preceding interval. There emerges the new
series of energies $ E_{(R)}^{(-\beta)} \ [\equiv b(N)]$ without
any left partners; this possibility represents just a weaker form
(and/or a more singular analogue) of the Jevicki-Rodriguez
breakdown of SUSY \cite{JR} as mentioned above in Abstract.

\item ``Central domain"
with $\gamma \in (-1,0)$. This is the most interesting domain
where the properties of the well known linear special case
$H^{(LHO)}$ (which has $\gamma = -1/2$) appear generalized to the
whole neighboring interval. Up to the exceptional (and newly
emerging) ground state $a(0)$ we may spot here the well known
pattern of degeneracy,
 % \newpage
 \ben
 E_{(L)}^{(-\alpha)} \ [\equiv a(N)] <
 E_{(R)}^{(-\beta)} \ [\equiv b(N)] =
 E_{(L)}^{(+\alpha)} \ [\equiv c(N)] <
 \een
 \ben
 < E_{(R)}^{(+\beta)} \ [\equiv d(N)] = a(N+1) < \ldots\
 \een
so that SUSY becomes unbroken even for the spectrum which ceased
to be equidistant at $\gamma \neq -1/2$.

\item ``Near right"
with $\gamma \in (0,1)$ and with the properties which ``mirror"
the far left (the series of energies $ E_{(R)}^{(-\beta)} \
[\equiv b(N)]$ ceases to exist, etc).

\item ``Far right"
with $\gamma \in (1, \infty)$, degeneracy
 \ben E_{(L)}^{(+\alpha)} \ [\equiv c(N)] = d(N-1), \ \ \ \
 N > 0
 \een
and with the characteristic $\gamma-$independence of the almost
completely degenerate unbroken SUSY spectrum.

\end{enumerate}

\noindent
We may summarize that the resulting SUSY pattern is fairly unusual.
It may be characterized by several above-listed appealing
properties but one should re-emphasize,
first of all, that
near $\gamma = -1/2$ a nice non-equidistant
generalization of the textbook $D=1$ SUSY oscillators is obtained.

\section*{Acknowledgement}

Work partially supported by the grant Nr. A 1048004 of GA AS CR.

\section*{Figure captions}

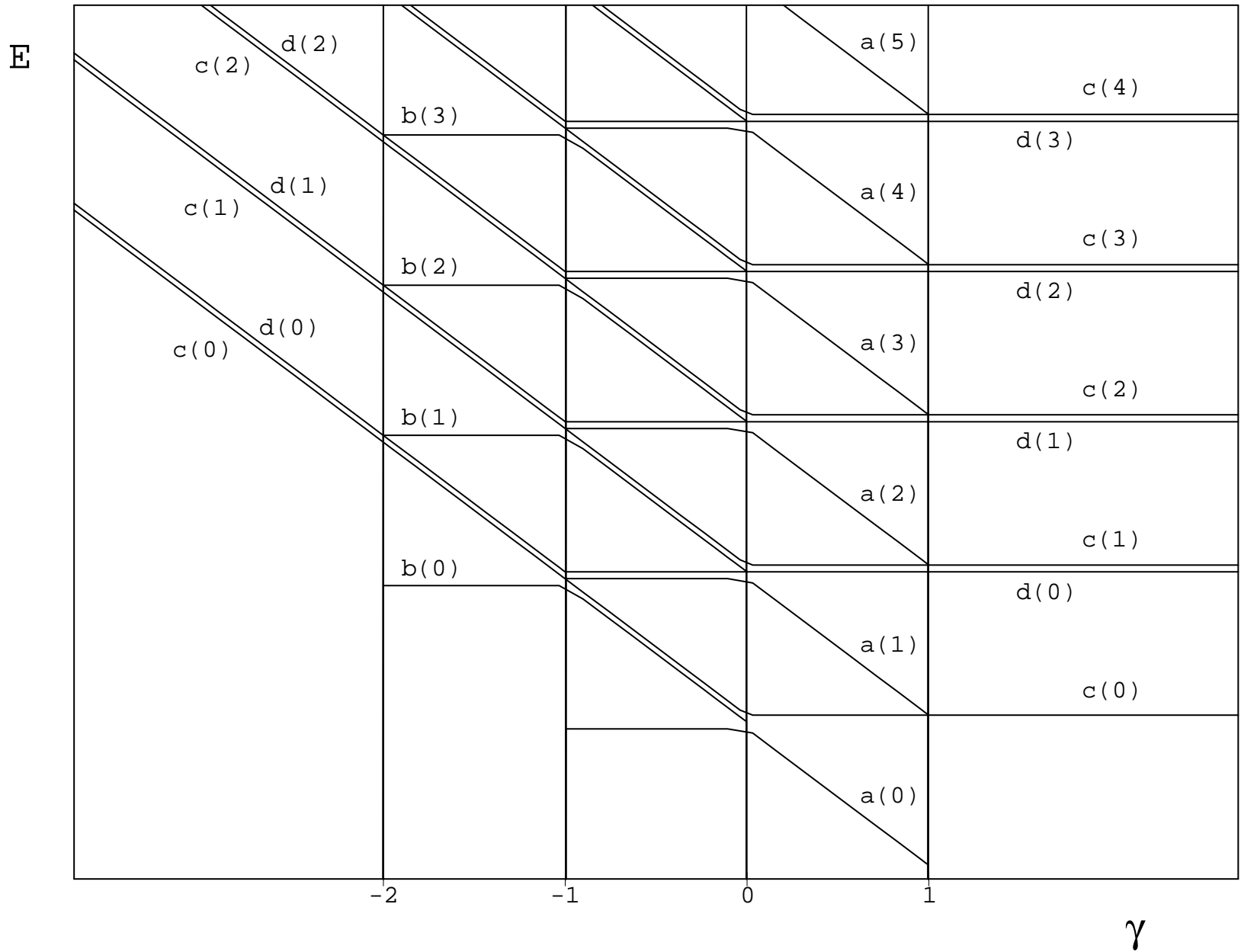
Figure 1. SUSY and the $\gamma-$dependence of the spectrum which
is generated by the superpotential~(\ref{N}).

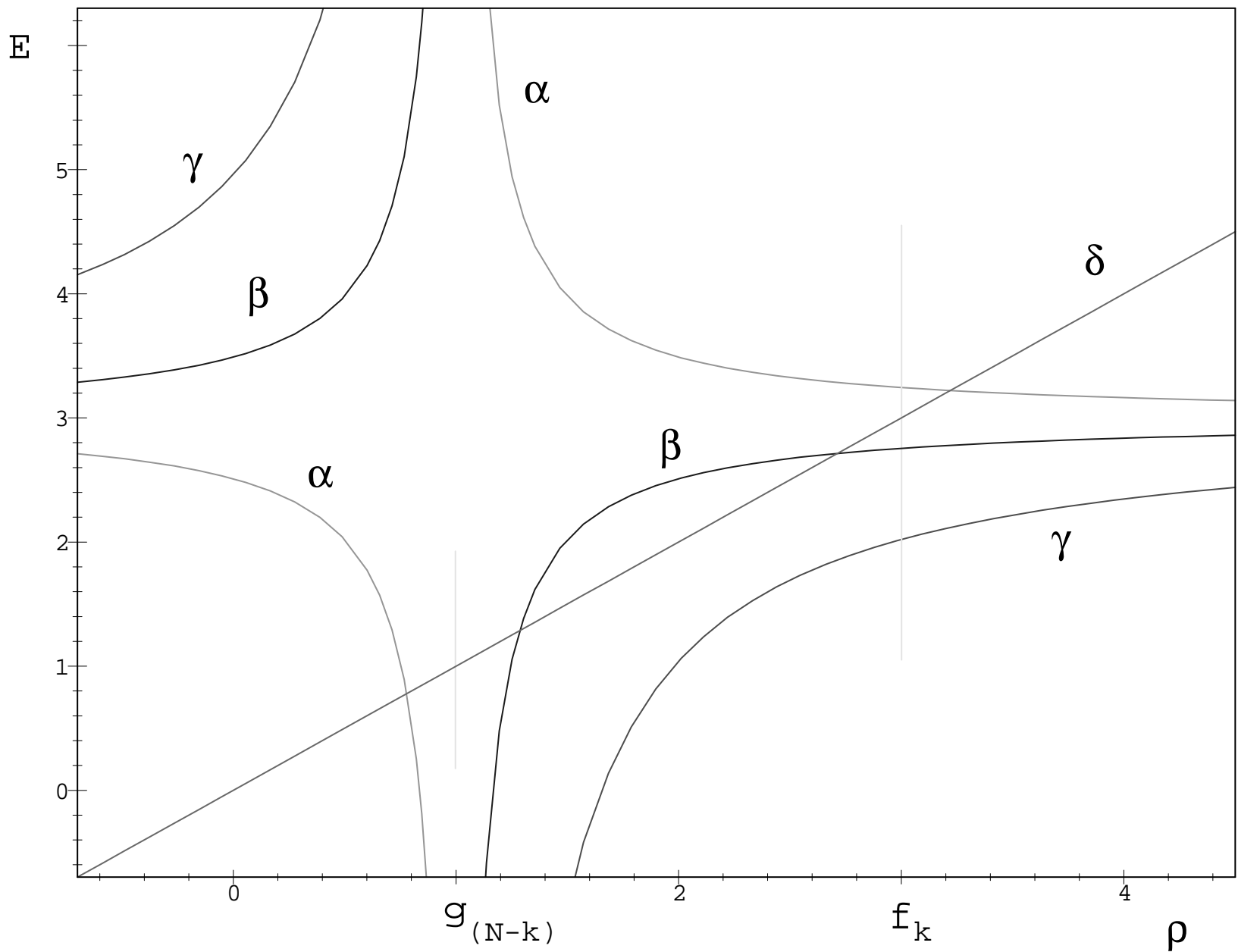
Figure 2. Graphical solution of the selfconsistency condition
$E(\varrho)=\varrho$ in the schematic example of Appendix B. 3.
with $g_{N-k}=1$ and $f_{k}=3$.

(A) curve (\ref{Efinro})
 at $a_k=0.7$ in the Hermitian regime.

(B) curve (\ref{Efinro}) at $a_k=0.7$ in the non-Hermitian regime
(both energies are real),

(C) curve (\ref{Efinro}) at $a_k=1.4$ in the non-Hermitian regime
(no real root, both energies are complex),

(D) selfconsistency line $E(\varrho)=\varrho$.

%\newpage

\section*{Appendix A: The standard harmonic oscillator basis
on $L_2(I\!\!R)$}

Eigenstates of a Hamiltonian $H(g)$ which commutes with the parity
${\cal P}$ may be numbered by an integer $n$ and by the
superscript $^\pm$ which characterizes the spatial parity of the
state,
 \be
 H(g)\,|n^{(\pm)}(g) \rangle = E_n^{(\pm)}(g)\,|n^{(\pm)}(g) \rangle
 \ee
These eigenstates form a complete basis in the Hilbert space
$L_2(I\!\!R)$,
 \be
\sum_{\sigma=\pm  }\ \sum_{m=0}^\infty\ |\,m^{(\sigma)}(g) \rangle
\,\langle m^{(\sigma)}(g)| = I\ .
 \label{complete}
 \ee
The usual condition of their orthonormalization reads
 \ben
   \langle n^{(\tau)}(g)|\,m^{(\sigma)}(g) \rangle =
\int_{-\infty}^\infty\
   \langle n^{(\tau)}(g)|\,x \rangle
   \langle x|\,m^{(\sigma)}(g) \rangle\ dx =
   \delta_{mn}\delta_{\sigma,\tau}.
 \label{fullline}
 \een
For the particular and exceptional harmonic oscillator $H(0)
\equiv H^{(LHO)}$ these eigenstates are proportional to the well
known Hermite polynomials,
 \be
 \ba
  \langle x|n^{(+)}(0) \rangle
=
{\cal N}_{2n} {\cal H}_{2n}(x)\,\exp(-x^2/2) \equiv
   \langle x|s_n \rangle,
   \\
 \langle x|n^{(-)}(0) \rangle
=
{\cal N}_{2n+1}
 {\cal H}_{2n+1}(x)\,\exp(-x^2/2) \equiv
   \langle x|t_n \rangle,\\ \ \ \ \ \ \ \ \ \ \ \ \ \ \ \
 \ x \in I\!\!R,
  \ \ \ \ \  {\cal N}_n=
\left ( \sqrt{2^n\,n!\,\sqrt{\pi}}
 \right )^{-1},\ \ \ \ n = 0, 1, \ldots
 \ .
 \ea
 \label{star}
 \ee
At each particular subscript $n=m$ the pairs of the latter
harmonic-oscillator basis states have an opposite parity, $ {\cal
P}\, |s\rangle =+|s\rangle, \ {\cal P}\, |t\rangle =-|t\rangle $.
They may be transformed into the unitarily equivalent pairs of
states
 \ben
 \left (
 \ba
 |S\rangle\\
 |T\rangle
 \ea
 \right ) =
 \frac{1}{\sqrt{2}}
 \,\left (
 \begin{array}{cc}
 1&i\\
 i&1
 \ea
 \right )
 \,
 \left (
 \ba
 |s\rangle\\
 |t\rangle
 \ea
 \right )
 \label{SppEr}
 \een
with the real ${\cal PT}$ parities, $ {\cal PT}\, |S\rangle
=+|S\rangle, \ {\cal PT}\, |T\rangle =-|T\rangle $ where the
complex conjugation ${\cal T}$ defined by the simple rule ${\cal
T}i{\cal T}=-i$ mimics the usual antilinear time reversal.

\section*{Appendix B: Main differences between the Hermitian and non-Hermitian
Hamiltonians}

\subsection*{B. 1: ${\cal P}-$symmetric models and the
bases on $L_2(I\!\!R^+)$}

Any eigenstate $|\psi\rangle$ of $H = H^\dagger ={\cal P}H{\cal
P}$ satisfies its Schr\"{o}dinger equation even after a
pre-multiplication by the parity ${\cal P}$. Both the old and new
eigenstates belong to the same real eigenvalue $E$ which cannot be
degenerate due to the Sturm-Liouville oscillation theorems. One of
the superpositions $|\psi\rangle \pm {\cal P}|\psi\rangle$ must
vanish while the other one acquires a definite parity.  This is
the essence of the mathematical proof that the ${\cal P}$ symmetry
of wave functions cannot be spontaneously broken, ${\cal P}
|n^{(\pm)}(g) \rangle= \pm |n^{(\pm)}(g) \rangle$.

The knowledge of the spatial symmetry of the Hermitian Hamiltonian
$H(g)$ enables us to simplify many considerations and calculations
by choosing and fixing the parity of the solutions in advance,
 \ben
\langle x|\,n^{(\pm)}(g)\rangle = \pm \langle
(-x)|\,n^{(\pm)}(g)\rangle.
 \een
This permits us to live, conveniently, on the semi-axis of $ x \in
(0, \infty) = I\!\!R^+$. In such a setting we rarely imagine that
we are {\em tacitly changing the Hilbert space} from $L_2(I\!\!R)$
to $L_2(I\!\!R^+)$. We feel that this is a technicality which
deserves a separate remark.

On the new space (or, if you wish, in the old space equipped by
the projector or singular metric $\Pi$) the inner product changes
its meaning since we have to integrate over the mere semi-axis
(symbolically, $\langle \psi|\psi'\rangle \to \langle
\psi|\Pi|\psi'\rangle$). This re-scales the orthogonality
relations,
 \ben
  \langle n^{(\sigma)}(g)|\,\Pi\, |\,m^{(\sigma)} (g)\rangle =
\int_{0}^\infty\
   \langle n^{(\sigma)}(g)|\,x \rangle
   \langle x|\,m^{(\sigma)}(g) \rangle\ dx =
\frac{1}{2} \delta_{mn}.
 \label{halfline}
 \een
Alternatively, we may omit the symbols $\Pi$ and switch to the two
re-normalized bases
 \ben
 \ba
 \langle x|\sigma_n \rangle
=
{\cal M}_{2n} {\cal H}_{2n}(x)\,\exp(-x^2/2)  ,
   \\
  \langle x|\tau_n \rangle
=
{\cal M}_{2n+1}
 {\cal H}_{2n+1}(x)\,\exp(-x^2/2)
 ,\\ \ \ \ \ \ \ \ \ \ \ \ \ \ \ \
 \ x \in I\!\!R^+,
  \ \ \ \ \  {\cal M}_n=
\left ( \sqrt{2^{n-1}\,n!\,\sqrt{\pi}}
 \right )^{-1},\ \ \ \ n = 0, 1, \ldots
 \
 \ea
 \label{starback}
 \een
which are both orthonormal on the half-line. In parallel,
condition (\ref{complete}) splits in the two independent
completeness relations
 \ben
 \sum_{m=0}^\infty\ |\,
 \sigma_m \rangle
\,\langle \sigma_m| = I,\ \ \ \ \ \ \ \  \sum_{n=0}^\infty\
|\,\tau_n \rangle \,\langle \tau_n| = I.
 \label{plete}
 \een
The overlaps of the states with different superscripts do not
vanish and form a unitary matrix which changes the basis in
$L_2(I\!\!R^+)$. Its elements
 \be
 {\cal U}_{n,m}=
 2\,\int_{0}^\infty\
   \langle n^{(+)}(0)|\,x \rangle
   \langle x|\,m^{(-)}(0) \rangle\ dx =
   2\,\langle s_n\,|\Pi \,|\,t_m \rangle=
   \langle \sigma_n\,|\,\tau_m \rangle
 \label{asyhalfline}
 \ee
may be computed by the direct symbolic integration in MAPLE giving
the exact values sampled in Table~1 from which one may extract
some closed formulae, e.g.,
 \ben
\langle s_n|\Pi |\,t_n \rangle = \frac{ (2n)! \,\sqrt{4n+2}}
{2^{2n+1}(n!)^2\, \sqrt{\pi}} =\frac{ \sqrt{n+\frac{1}{2}}}{\pi}
\,\frac{ \Gamma(n+\frac{1}{2}) } { \Gamma(n+{1}) }.
 \een
This means that the original vectors (\ref{star}) form an
over-complete set and we may make a choice between the two
alternative basis sets $\{|\sigma\rangle\}$ and
$\{|\tau\rangle\}$. They are both complete on the new Hilbert
space $L_2(I\!\!R^+)$.

\subsection*{B. 2: ${\cal PT}-$symmetric models }

The above-mentioned rigidity of the conservation of parity is lost
during the transition to the ${\cal PT}$ symmetric models  $H =
H^\ddagger ={\cal PT}H{\cal PT}$  where any quantity
$\exp(i\varphi)$ is an admissible eigenvalue of the operator
${\cal PT}$  since its component ${\cal T}$ is defined as
anti-linear, ${\cal T}i=-i$. In more detail, every rule ${\cal
PT}|\psi\rangle = \exp(i\varphi)\,|\psi\rangle$ implies that we
have
 \ben
 {\cal PT}{\cal PT}|\psi\rangle =
\exp(-i\varphi)\,{\cal PT}|\psi\rangle= |\psi\rangle
 \een
as required. The Schur's lemma ceases to be applicable. In the
basis of Appendix A with the properties ${\cal PT} |S\rangle =
|S\rangle$ and ${\cal PT} |L\rangle = -|L\rangle$, the general
expansion formula
 \ben
 H= \sum_{m,n=0}^{\infty}  \left (
 \ba
 \\
 \ea
 |S_m\rangle {\cal F}_{m,n} \langle S_n | \,+\,
 |L_m\rangle {\cal G}_{m,n} \langle L_n | \,+\,
 i\,|S_m\rangle {\cal  C}_{m,n} \langle L_n | \,+\,
 i\,|L_m\rangle {\cal D}_{m,n} \langle S_n |\
 \right )
 \een
contains four separate complex matrices of coefficients.  Once it
is subdued to the requirement $H= {\cal PT} H{\cal PT}$, we get
the necessary and sufficient condition demanding that all the
above matrix elements of $H=H^\ddagger$ must be real,
 \be
 {\cal F}_{m,n} = {\cal F}_{m,n}^*, \ \ \ \ \  {\cal G}_{m,n}
= {\cal G}_{m,n}^*, \ \ \ \ \   {\cal C}_{m,n} = {\cal C}_{m,n}^*,
\ \ \ \ \ {\cal D}_{m,n} = {\cal D}_{m,n}^*.
 \label{constrainte}
 \ee
As long as the similar trick has led to the superselection rules
for the spatially symmetric Hamiltonians, we may conclude that the
${\cal PT}$ symmetric analogue of the direct-sum decompositions
and superselection rule is just the much weaker
constraint~(\ref{constrainte}).

\subsection*{B. 3: Schematic finite-dimensional matrix model
with and without ${\cal PT}$ symmetry}

Let us contemplate the partitioned matrix Schr\"{o}dinger equation
 \be
\left (
\begin{array}{cc}
F-E\,I&\alpha\,A\\ A^\dagger& G-E\,I
 \ea
\right ) \cdot \left ( \ba \vec{u}\\ \vec{w}
 \ea
\right ) = 0\ \label{SEdrd}
 \ee
where $F = F^dagger$, $G = G^\dagger$ and either $\alpha=1$
(Hermitian case) or $\alpha=-1$ (${\cal PT}$ symmetric case).
Schr\"{o}dinger equations with the matrix representation
(\ref{SEdrd}) generalize the models with ${\cal PT}$ symmetry
\cite{Mostafa2}. Their spectrum may happen to be real and
discrete, at least in the limit $A \to 0$, or containing the
complex conjugate pairs. Let us now descibe their nontrivial,
non-perturbative solvable example.

Preliminarily, both the Hermitian submatrices $F$ and $G$ of the
Hamiltonian should be diagonalized via a pair of some suitable
unitary transformations, $F\to \hat{f}$,  $G \to \hat{g}$. Their
respective spectra $\{f_n\}$ and $\{g_n\}$ will be assumed real
and discrete.

Secondly, we shall ignore all the small elements of the coupling
matrix ${\cal A}$ in our pre-diagonalized effective
Schr\"{o}dinger equation (\ref{SEdrd}),
 \be
 \left (
\hat{f}^{}-E\,I- \alpha\, {\cal A}\,
 \frac{1}{\hat{g}^{}-E\,I}\,
 {\cal A}^\dagger
 \right )
\vec{y} = 0.
 \label{pStEfin}
 \ee
Here, $\alpha = \pm 1$ ``remembers" its respective Hermitian and
non-Hermitian origin and all the small elements in ${\cal A}$ are
irrelevant causing just a perturbative, small deformation of the
decoupled spectrum $\{f_n\} + \{g_n\}$.

Thirdly, let us choose the latter coupling matrix in the form
dominated by the transposed diagonal,
 \ben
 {\cal A} =
 \left (
 \begin{array}{ccccc}
0&0&\ldots&0&a_0\\ 0&\ldots&0&a_1&0\\ \vdots&&&\vdots&\vdots\\
0&a_{N-1}&0&\ldots&0\\ a_{N}&0&\ldots&0&0
 \ea
 \right ).
 \een
Quite unexpectedly, this choice makes the problem exactly solvable
since the secular equation $\det (H_{eff}(\varrho)-EI)=0$ can be
immediately factorized,
 \ben
0 = \left ( f_0-E-\alpha \frac{|a_0|^2}{g_N-\varrho} \right )
\cdot \left ( f_1-E-\alpha \frac{|a_1|^2}{g_{N-1}-\varrho} \right
) \cdots \ldots \cdot \left ( f_N-E-\alpha
\frac{|a_N|^2}{g_{0}-\varrho} \right ).
 \een
The explicit evaluation of zeros of the $k-$th factor is trivial,
 \be
E(\varrho)={f}^{}_k- \alpha\, a_k\,
 \frac{1}{{g}^{}_{N-k}-\varrho}\,
a^*_k.
 \label{Efinro}
 \ee
The implementation of the selfconsistency $\varrho=E(\varrho)$
gives the sequence of the
 mere quadratic algebraic equations
 \ben
{f}^{}_k-E- \alpha\, a_k\,
 \frac{1}{{g}^{}_{N-k}-E}\,
a^*_k = 0.
 \label{Efin}
 \een
All their roots are available in closed form,
 \ben
 E_{k^\pm}
= \frac{1}{2} \left (
f_k+g_{N-k}\pm\sqrt{(f_k-g_{N-k})^2+4\alpha\,|a_k|^2} \right ).
 \een
This confirms our {\it a priori} expectations since the Hermitian
energies with $\alpha=1$ are always real while, at $\alpha = -1$,
we get the real spectrum if and only if
 \be
|a_k| < |f_k-g_{N-k}|/2, \ \ \ \ k = 0, 1, \ldots, N.
\label{strongs}
 \ee
Vice versa, we get a complex conjugate pair $ E_{k^\pm} $ whenever
we move to the strongly non-Hermitian regime and encounter a large
and strong off-diagonality or coupling of modes in ${\cal A}$.
This is an independent linear-algebraic re-confirmation of the
similar observations made during the explicit computations using
differential equations.

\newpage

Table 1. Overlaps $\sqrt{\pi}\, \langle s_n|\Pi |\,t_m \rangle $
defined by eq. (\ref{asyhalfline}). Rows are numbered by
 $n=0,1,\ldots,6$, columns by
 $m=0,1,\ldots,4$.

 $$
 \begin{array}{||c|c|c|c|c||}
 \hline \hline

1/2\,\sqrt {2}
                                       &

-1/6\,\sqrt {3}
                                       &

1/20\,\sqrt {15}
                                       &

-{\frac {1}{56}}\,\sqrt {70}
                                       &

1/48\,\sqrt {35}

                                        \\ \hline
 1/2
                                       &

1/4\,\sqrt {3}\sqrt {2}
                                       &

-1/24\,\sqrt {15}\sqrt {2}
                                       &

{\frac {1}{80}}\,\sqrt {70}\sqrt {2}
                                       &

-{\frac {3}{224}}\,\sqrt {35}\sqrt {2}

                                       \\ \hline

-1/24\,\sqrt {2}\sqrt {6}
                                       &

1/8\,\sqrt {3}\sqrt {6}
                                       &

1/16\,\sqrt {15}\sqrt {6}
                                       &

-{\frac {1}{96}}\,\sqrt {70}\sqrt {6}
                                       &

{\frac {3}{320}}\,\sqrt {35}\sqrt {6}

                                       \\ \hline

1/40\,\sqrt {2}\sqrt {5}
                                       &

-1/24\,\sqrt {3}\sqrt {5}
                                       &

1/16\,\sqrt {15}\sqrt {5}
                                       &

1/32\,\sqrt {70}\sqrt {5}
                                       &

-{\frac {1}{64}}\,\sqrt {35}\sqrt {5}

                                       \\ \hline

-{\frac {1}{224}}\,\sqrt {70}\sqrt {2}
                                       &

{\frac {1}{160}}\,\sqrt {3}\sqrt {70}
                                       &

-{\frac {1}{192}}\,\sqrt {15}\sqrt {70}
                                       &

{\frac {35}{64}}
                                       &

{\frac {3}{256}}\,\sqrt {35}\sqrt {70}

                                       \\ \hline

{\frac {1}{96}}\,\sqrt {2}\sqrt {7}
                                       &

-{\frac {3}{224}}\,\sqrt {3}\sqrt {7}
                                       &

{\frac {3}{320}}\,\sqrt {15}\sqrt {7}
                                       &

-{\frac {1}{128}}\,\sqrt {70}\sqrt {7}
                                       &

{\frac {9}{256}}\,\sqrt {35}\sqrt {7}

                                        \\ \hline

-{\frac {1}{704}}\,\sqrt {2}\sqrt {231}
                                       &

{\frac {1}{576}}\,\sqrt {3}\sqrt {231}
                                       &

-{\frac {1}{896}}\,\sqrt {15}\sqrt {231}
                                       &

{\frac {1}{1280}}\,\sqrt {70}\sqrt {231}
                                       &

-{\frac {1}{512}}\,\sqrt {35}\sqrt {231}

                                       \\
 \hline
\hline
 \end{array}
  $$

\end{document}